\documentclass[10pt,a4paper]{article}
\usepackage{amsmath}
\usepackage{amsfonts}
\usepackage{amssymb}
\usepackage{graphicx}
\usepackage{textcomp}
\usepackage{url}
\usepackage{lipsum}
\usepackage{setspace}
\usepackage{color}
\usepackage[numbers,sort&compress]{natbib}
\usepackage[usenames,dvipsnames,svgnames,table]{xcolor}
\usepackage[lofdepth,lotdepth,caption=false]{subfig}
\usepackage[pdfstartview=FitH]{hyperref}
\usepackage{hyperref}
\newsavebox{\bigleftbox}
\hypersetup{
 colorlinks,
 citecolor=Blue,
 linkcolor=Blue,
 urlcolor=Blue
}

\makeatletter
 \def\footnoterule{\kern-3\p@
   \noindent\hrulefill \kern 2.8\p@} 
\makeatother
\usepackage[left=3.00cm, right=3.00cm, top=2.00cm, bottom=2.00cm]{geometry}

\usepackage{fancyhdr}

\title{\textbf{
Irida-Graphene: A New 2D Carbon Allotrope
}}
\author{
	M. L. Pereira Júnior$^{1,\ddag}$,
	W. F. da Cunha$^{2}$,
	W. F. Giozza$^{1}$,\\
	R. T. de Sousa Junior$^{1}$, and
	L. A. Ribeiro Junior$^{2,\dag}$
	}
\date{}

\begin{document}
    \maketitle
	\vspace{-0.6cm}
	\begin{center}\small
		$^1$\textit{Department of Electrical Engineering, Faculty of Technology, University of Bras\'{i}lia, Bras\'{i}lia, Brazil}\\
		$^2$\textit{Institute of Physics, University of Brasília, Brasília, Brazil}\\ \vspace{0.2cm} 
		\phantom{.}\hfill
		$^{\ddag}$\url{marcelo.lopes@unb.br}\hfill
		$^{\dag}$\url{ribeirojr@unb.br}\hfill
		\phantom{.}
	\end{center}
	

\onehalfspace

\noindent\textbf{Abstract:} Several 2D carbon-based materials have been computationally designed in the last years due to the success achieved by graphene. Here, we propose a new 2D all-sp$^2$ carbon allotrope, named Irida-Graphene (IG), using a bottom-up approach. IG is composed of fused rings containing 3-6-8 carbon atoms. We employed density functional theory calculations and reactive (ReaxFF) molecular dynamics simulations to examine its mechanical, structural, electronic, and optical properties. Results showed that IG exhibits good dynamical and thermal stabilities. Its estimated elastic modulus varies between 80-113 GPa. IG is a metallic material and presents a Dirac cone above the Fermi level in the center of the band. The intense optical activity of IG is restricted to the infrared and violet regions. IG can act as a violet collector for photon energies of about 3.0 eV since it presents very low reflectivity and refractive index greater than one.   

\section{Introduction}

The design \cite{yang2013theoretical,zhang2019art} and synthesis \cite{toh2020synthesis,fan2021biphenylene,hou2022synthesis} of new carbon allotropes have experienced enormous growth since the advent of graphene \cite{geim2009graphene,geim2010rise}. Its 2D honeycomb arrangement with an atomic thickness exhibits unique mechanical, optical, and electronic properties, which are of significant interest for many potential applications in nanoelectronics \cite{westervelt2008graphene}. Due to this reason, several 2D carbon-based materials have been computationally proposed in the last years \cite{enyashin2011graphene,xu2014two,terrones2000new,paz2019naphthylenes,lu2013two,wang2015phagraphene,zhang2015penta,wang2018popgraphene,jana2021emerging}. Among them, the monolayer amorphous carbon \cite{toh2020synthesis}, biphenylene network \cite{fan2021biphenylene}, and monolayer fullerene network \cite{hou2022synthesis} were synthesized recently. The successful synthesis of these materials has helped to stimulate the designing of new 2D carbon allotropes that can be obtained experimentally. 

A trend in developing new carbon allotropes has been to propose structures with non-hexagonal rings and large pores. The ultimate motivation is that carbon-based materials with non-hexagonal ring structures are better adsorber of Lithium atoms when contrasted with graphene \cite{yu2013graphenylene}. One can highlight the popgraphene (composed of 5-8-5 carbon rings) \cite{wang2018popgraphene}, phagraphene \cite{wang2015phagraphene}, and psi-graphene \cite{li2017psi}, which are composed of 5-6-7 carbon rings. They are intrinsically metallic and have a high theoretical capacity to adsorb Li atoms (1487, 556, and 372 mA h g$^{-1}$, respectively), which are crucial traits for developing efficient Li-ion batteries. Based on these performances for Li storage capacity, it can be helpful in energy storage applications to design other porous 2D metallic materials containing fused rings with eight carbon atoms. 

In this study, we employed a bottom-up approach to computationally propose a new 2D all-sp$^{2}$ carbon allotrope composed of 3-6-8 rings, named Irida-Graphene (IG). This name is due to its atomic arrangement that resembles the flower (\textit{Sisyrinchium Halophilum - Iridaceae}) popularly called "Nevada Blue-Eyed Grass" in United States (see Figure \ref{fig1}). We carried out density functional theory (DFT) and reactive (ReaxFF) molecular dynamics (MD) simulations to study the electronic, optical, and mechanical properties of this material. DFT and MD results revealed that IG is metallic and structurally stable, as confirmed by its integrity at 4176 K and by the absence of imaginary phonon modes in its phonon dispersion spectra. Its optical activity is restricted to the infrared and violet regions.              

\section{Methodology}

\begin{figure*}[!htb]
	\centering
	\includegraphics[width=1.0\linewidth]{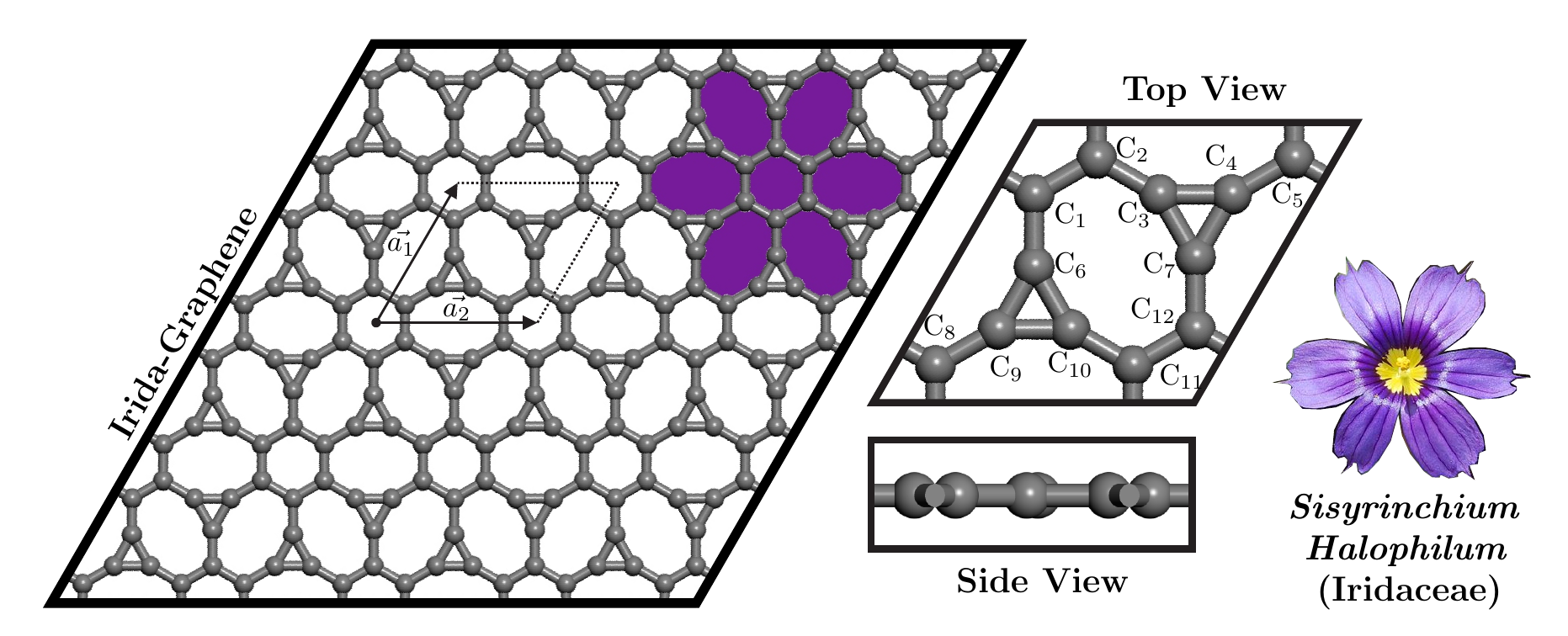}
	\caption{Schematic representation of the Irida-Graphene structure. The right panels show the top and side views of its unit cell and a representation of the \textit{Sisyrinchium Halophilum - Iridaceae} flower.}
	\label{fig1}
\end{figure*}

The lattice structure of IG is shown in Figure \ref{fig1}. Its unit cell is flat and composed of 12 carbon atoms. As one can see in this figure, IG is a 2D all-sp$^{2}$ carbon allotrope composed of fused 3-6-8 rings. There are three types of carbon-carbon bonds in its structure: C$_1$-C$_2$/C$_{11}$-C$_{12}$/C$_{1}$-C$_{5}$/C$_{2}$-C$_{8}$/C$_{2}$-C$_{8}$/C$_{8}$-C$_{11}$ = 1.439 \r{A},  C$_2$-C$_3$/C$_4$-C$_5$/C$_1$-C$_6$/C$_{7}$-C$_{12}$/C$_{8}$-C$_{9}$/C$_{10}$-C$_{11}$ = 1.410 \r{A}, and  C$_3$-C$_4$/C$_3$-C$_7$/C$_4$-C$_7$/C$_9$-C$_9$/C$_{6}$-C$_{10}$/C$_{9}$-C$_{10}$ =  1.413 \r{A}. 

\subsection{First-Principles Calculations}

All first-principles calculations were based on DFT theory as implemented in the SIESTA code \cite{Soler2002,artacho2008siesta,artacho1999linear}. These calculations were carried out within the scope of the generalized gradient approximation (GGA) with the Perdew-Burke-Ernzerhof (PBE) \cite{perdew1996generalized,ernzerhof1999assessment} functional for the exchange-correlation term. Norm-conserving Troullier-Martins pseudopotential \cite{troullier1991efficient} was employed to describe the core electrons. The wave functions were described by localized atomic orbitals and a double-zeta plus polarization (DZP) basis set. 

We used 500 Ry for kinetic energy cut-off, sampling the reciprocal space with 30 $\times$ 30 $\times$ 1 k-point grid. The vacuum region along the z-direction is set to 20 \r{A} to prevent spurious interactions between the sheet and its periodic images. The basis vector along the z-direction was fixed during the lattice optimization. The other vectors and atoms were fully relaxed. The convergence criterion for maximum forces on each atom was 0.001 eV/\r{A}. To investigate the stability of IG, we performed phonon dispersion calculations based on the force constant algorithm for T${}=0$ K, as implemented in Phonopy \cite{togo2015first}, and Born-Oppenheimer \textit{ab initio} molecular dynamics simulations (AIMD) for 300 K with the canonical ensemble NVT. For the electronic band structure calculation, we considered the following path along the lattice: $\Gamma = (0,0)$ to $K = (0,1/2)$ to $Y = (1/3,1/3)$ to $Z=(1/2,0)$ to $\Gamma = (0,0)$. 

To estimate the optical coefficients, an external electric field of value 1.0 V/\r{A} was applied along three possible polarization planes. The optical coefficients were extracted from complex dielectric function $\epsilon =\epsilon_1+i\epsilon_2$, in which the imaginary part, $\epsilon_2$, is derived from the direct interband transitions through Fermi's golden rule. A detailed description on the procedure to calculate of the optical properties (used in this work) is given in reference \cite{tromer2021optoelectronic}. 

\subsection{Reactive (ReaxFF) Molecular Dynamics Simulations}

The mechanical and thermal properties of IG were investigated by employing fully atomistic MD simulations with the reactive force field ReaxFF \cite{senftle2016reaxff}, as implemented in LAMMPS \cite{thompson2022lammps,plimpton1995fast}. The ReaxFF potential allows the formation and breaking of chemical bonds during the dynamics, which is crucial to analyzing the fracture mechanisms of the materials. We used the ReaxFF parameter set for C/H/O \cite{smith2017reaxff}. The equations of motion were numerically integrated using the velocity-Verlet algorithm with a time-step of $0.05$ fs. The uniaxial tensile loading was applied along the non-periodic $x$ and $y$ directions, for an engineering strain rate of $1.0 \times 10^{-6}$ fs$^{-1}$. Before the stretching of the material, the lattice was equilibrated using an NPT ensemble at a constant temperature of 300 K and null pressures for a Nos\'e-Hoover thermostat \cite{hoover1985canonical} during 100 ps. 

To calculate the mechanical properties, the IG structure was continuously stretched up to its complete fracture. The maximum applied strain was 50\%. We obtained the following elastic properties from the stress-strain curves: Young’s modulus ($Y_M$), Fracture Strain (FS), and Ultimate Strength (US). The length in the z-direction of the simulation box for the mechanical stretching and heating simulations is 20 nm to prevent spurious interactions between the sheet and its periodic images. 

The thermal stability of IG was simulated in a vacuum by heating the simulation box from 300 up to 10000 K under periodic boundary conditions. Before heating, the structure was thermalized and equilibrated within an NPT ensemble to eliminate any residual thermal stresses and adjust the simulation box dimensions in the xy-plane. The heating process is simulated by a linear temperature increase during 2 ns in an NVT ensemble, defining a temperature increase velocity of $\sim$ 197 K/ms. 

We calculated the von Mises stress (VM) per-atom values (see Supplementary Material). These values provide information on the fracture process once they reveal the fracture point or region. MD snapshots illustrating the fracture patterns and the melting process of IG are shown in the Supplementary Material. The MD snapshots and trajectories were obtained using the visualization and analysis software VMD \cite{HUMPHREY199633}.

\section{Results}

The discussion is divided into three subsections: I) structural stability and mechanical properties and II) electronic and optical properties. The Supplementary Material presents the MD snapshots for the uniaxial tensile loading and the heating of IG.   

\subsection{Structural Stability and Mechanical Properties}

\begin{figure}[!htb]
	\centering
	\includegraphics[width=0.5\linewidth]{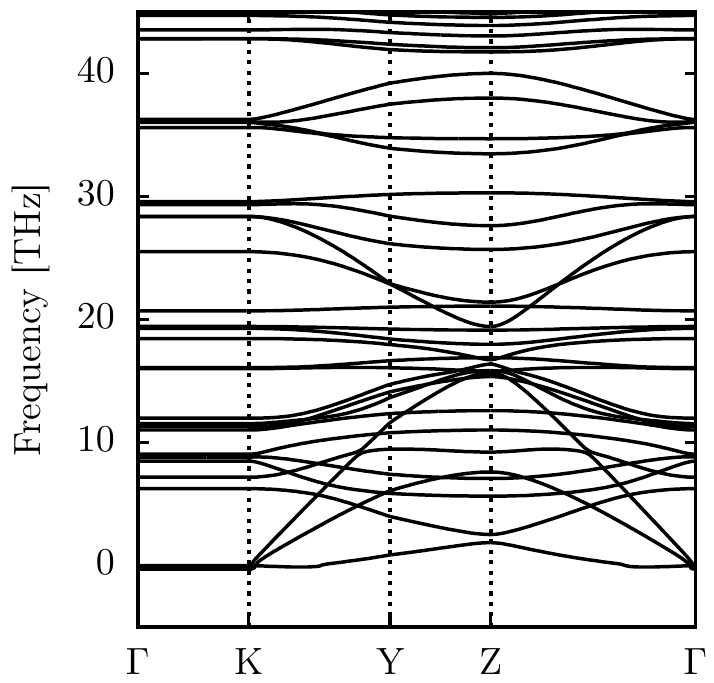}
	\caption{Phonon band structure of Irida-Graphene with a $2\times2\times1$ supercell.}
	\label{phonon}
\end{figure}

\begin{figure}[!htb]
	\centering
	\includegraphics[width=0.7\linewidth]{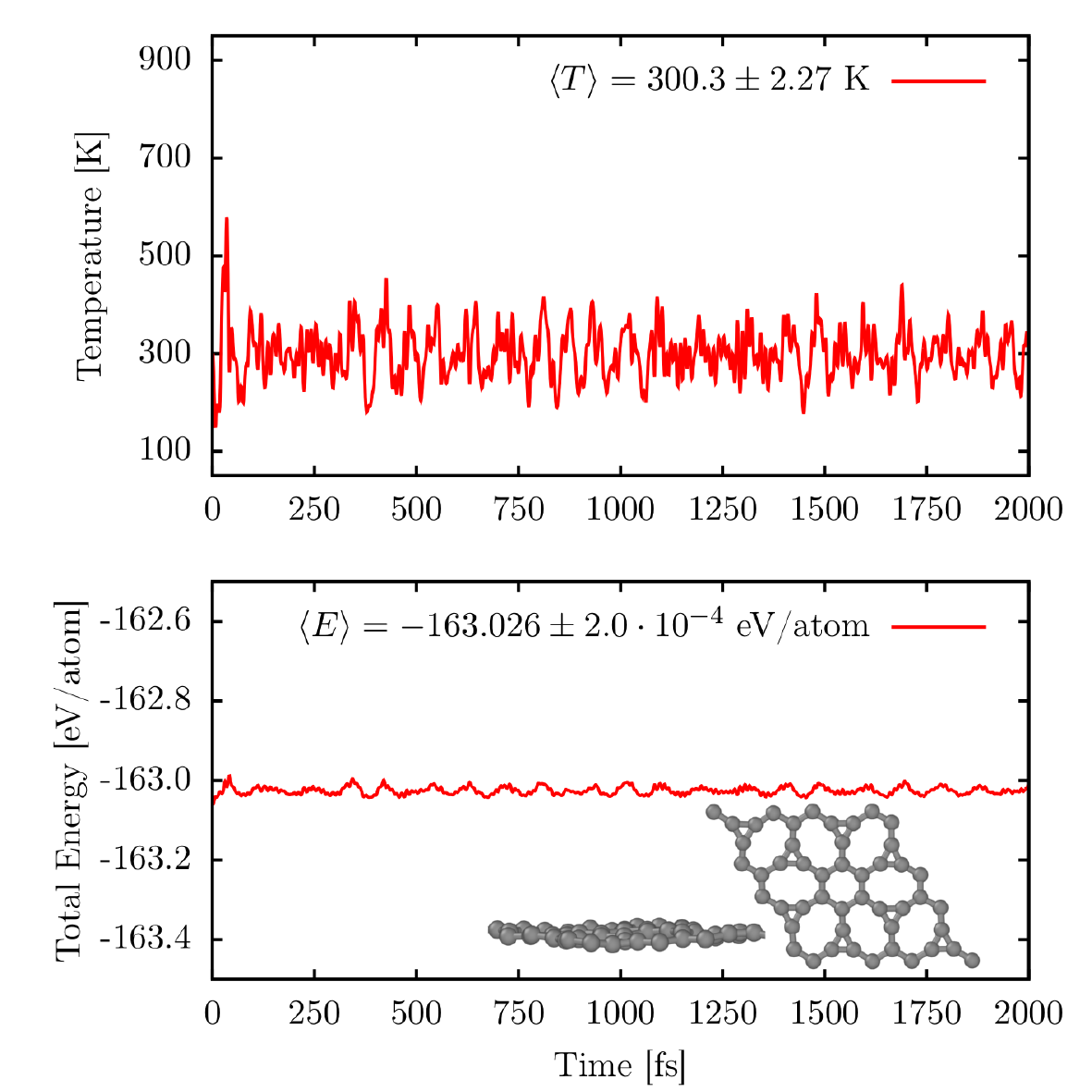}
	\caption{Time evolution of the total energy per atom in the IG lattice. The insets show the top and side views of the final AIMD snapshot at 2 ps.}
	\label{aimd}
\end{figure}

We examined the dynamical stability of IG by calculating the phonon dispersion relations along the high symmetry directions shown in Figure \ref{phonon}. Furthermore, we carried out 2 ps of AIMD at 300 K to ensure the IG stability, as depicted in Figure \ref{aimd}. In Figure \ref{phonon}, one can see that all phonon frequencies are positive without any imaginary modes in the Brillouin zone. This trend for the phonon spectrum confirms the IG dynamical and structural stability. Moreover, the highest phonon frequency is about 47 THz, similar to graphene \cite{song2018two} and indicating the presence of only sp$^{2}$ carbon-carbon bonds.                 

Figure \ref{aimd} shows the results for the AIMD simulation at 300 K. The insets illustrate the top and side views for the AIMD snapshot of the IG lattice in the final stage of the simulation. We adopted a 2$\times$2$\times$1 supercell containing 64 carbon atoms. One can note that the total energy per atom for the IG lattice fluctuates around a steady level in the heating process (equilibrium state). Moreover, in the MD snapshot for IG at 2 ps, the lattice structure maintains its planarity and has no bond reconstructions. The features revealed in the AIMD simulations suggest that IG has good thermal stability.                

\begin{figure}[!htb]
\centering
\includegraphics[width=0.5\linewidth]{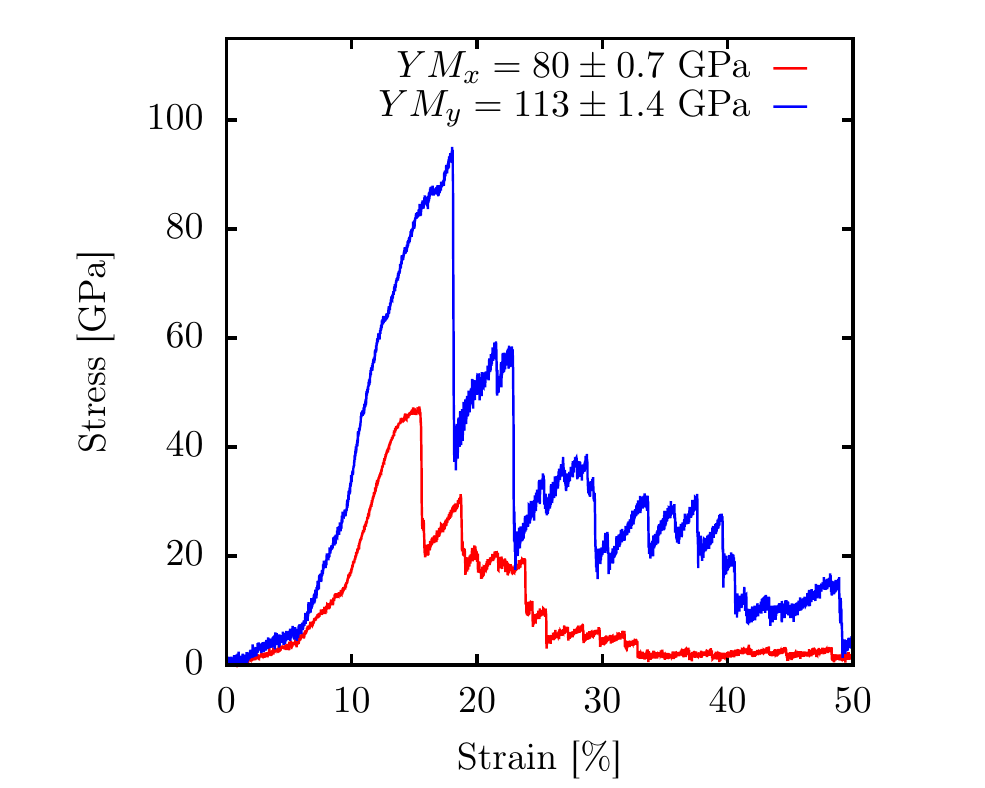}
\caption{Stress-strain relationships of IG under uniaxial tensile loading in (red) the x-direction and (blue) y-direction.}
\label{stress-strain}
\end{figure}

The mechanical properties of IG are studied through reactive ReaxFF MD simulations. The IG lattice has periodic boundary conditions and is subjected to a uniaxial tensile strain between 0-50\%. Figure \ref{stress-strain} presents the stress-strain curve under uniaxial strain loading along the $x$ and $y$ directions (black and blue lines, respectively). The Young's modulus values $YM_{X}$ and $YM_{Y}$ (along the x- and y-directions, respectively) are calculated considering the first linear regime, up to 1\% of strain for each curve. As a general trend, there are two well-defined regimes for the stress-strain relationship in IG lattices. 

In the first regime, between 0-15\% and 14\% of strain in the x- and y-direction, respectively, the stress rises sharply. In these regimes, the IG lattice keeps its integrity, but the carbon rings suffer deformations during the stretching process (see Supplementary Material). At 16\% and 15\% of strain for the tensile loading in x- and y-direction, respectively, the IG lattice suffers its first bond braking, producing a crack, which propagates rapidly through the structure. Above these critical strain thresholds (second regime for the stress-strain relationship), the stress drops abruptly, indicating the rupture of the IG lattice. The fluctuations in the curves shown in Figure \ref{stress-strain} denote several formations of linear atomic chains (LACs) during the stretching process (see Supplementary Material).

The estimated elastic moduli are $YM_{X}=80$ GPa and $YM_{Y}=113$ GPa. These results revealed that IG has clear in-plane anisotropic properties, as expected from the structure topology. The eight-membered rings of carbon atoms are responsible for the smaller mechanical resilience presented by IG when compared with graphene (about 1 TPa) \cite{lee2008measurement}. However, the $YM$ values calculated for IG are similar to those reported for other graphene-like allotropes \cite{sui2017morphology,sun2016new}. The IG fracture strain ($FS$) and ultimate stress ($US$) are: $FS=15.4\%$ and $US=47.4$ GPa in the x-direction and $FS=18.0\%$ and $US=95.1$ GPa in the y-direction. It is worth mentioning that FS is obtained from the strain percentage corresponding to the largest stress value, i.e., the US. These mechanical properties are also similar to those for other graphene-like allotropes \cite{sui2017morphology,sun2016new}. 

\begin{figure}[!htb]
    \centering
	\includegraphics[width=0.5\linewidth]{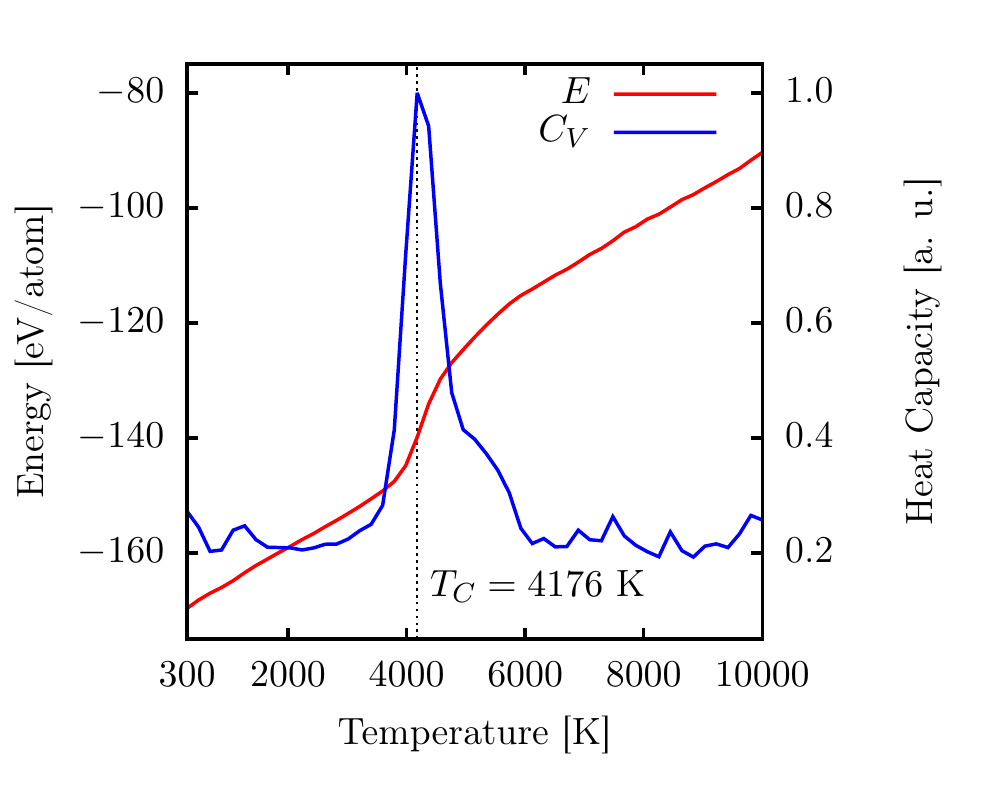}
	\caption{Irida-graphene total energy and heat capacity ($C_V$) as a function of temperature for the 250 ps heating ramp simulation.}
	\label{melting}
\end{figure}

We further explore the thermal stability of IG by studying its melting process through ReaxFF MD simulations. In our heating ramp protocol, the temperatures vary from 300 K up to 10000 K during 250 ps. Figure \ref{melting} shows the total energy (red) and heat capacity (blue) as a function of temperature for the melting process of IG. MD snapshots for the IG melting process are presented in the Supplementary Material. During the IG heating, the total energy increases quasi-linearly with the temperature with three different regimes defined by distinct slopes in the red curve, as shown in Figure \ref{melting}: between 300K-5000K, 5000K-5500K, and 5500K-10000K. 

In the first heating regime (300K-4000K), IG maintains its structural integrity. Beyond 4000K, the thermal vibrations lead to considerable changes in the original IG morphology. This structural phase transition is characterized by the first peak and discontinuity in the $C_V$ and total energy curves, respectively. 

The second heating regime (between 4000K-5500K) defined the IG transition from the solid to gas phase (melting process). The IG melting is denoted by the well-pronounced peak in the $C_V$ curve and a discontinuity in the total energy curve. The melting point occurs at 4176K. Importantly, this value is comparable to the melting point for the monolayer graphene (4095K) \cite{felix2020mechanical}. The complete IG melting takes place at 5500K. For higher temperatures, the total energy curve does not change the slope. 

\subsection{Electronic and Optical Properties}

\begin{figure}
	\centering
	\includegraphics[width=0.7\linewidth]{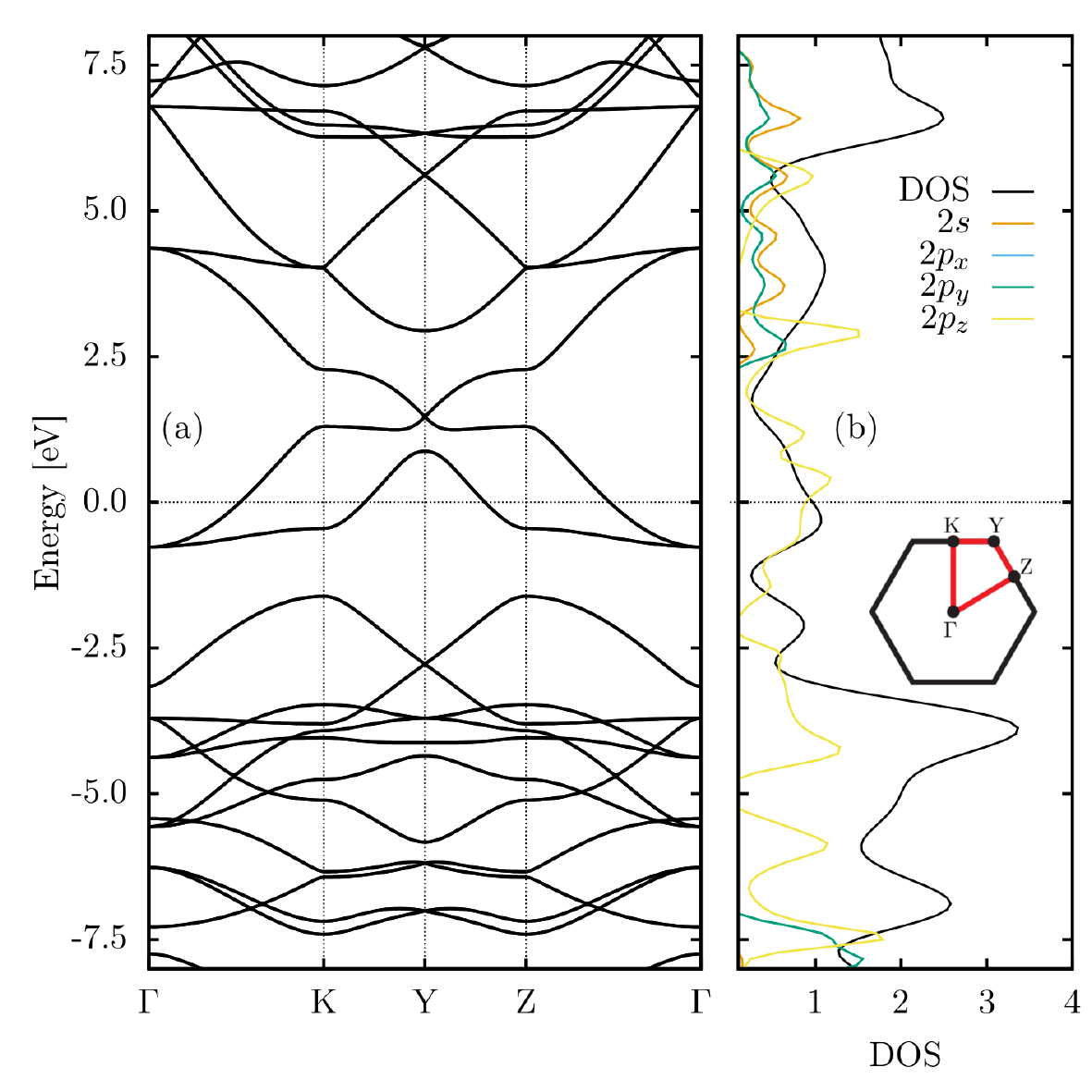}
	\caption{Electronic band structure and related partial density of states (PDOS) of IG. We considered the following path along the high symmetry directions: $\Gamma = (0,0)$ to $K = (0,1/2)$ to $Y = (1/3,1/3)$ to $Z=(1/2,0)$ to $\Gamma = (0,0)$.} 
	\label{bands}
\end{figure}

The calculated band structure, related density of states (DOS), and partial density of states (PDOS) are shown in Figure \ref{bands}. The inset in Figure \ref{bands}(a) illustrates the high symmetry directions used to calculate the electronic band structure. On the PBE level, no band gap was found between the valence and conduction bands (see Figure \ref{bands}), indicating that IG is metallic. The DOS near the Fermi level is essentially formed by 2p$_z$ atomic orbitals, confirming its metallic signature. 

IG valence bands are predominantly composed of 2p$_z$ atomic orbitals, with a small contribution of 2p$_y$ orbitals. On the other hand, the conduction bands are formed by also by 2s, 2p$_y$, and 2p$_z$ orbitals with higher contributions. This trend for the higher contribution of 2p$_z$ orbitals to the conduction bands is typical in 2D carbon-based systems \cite{gui2008band}. A Dirac cone is situated above the Fermi level and in the center of the band at the $Y$-point. Other 2D materials also present such a trend for the localization of the Dirac cone \cite{xu2014two,wang2015phagraphene}. 

\begin{figure}[!htb]
	\centering
	\includegraphics[width=0.5\linewidth]{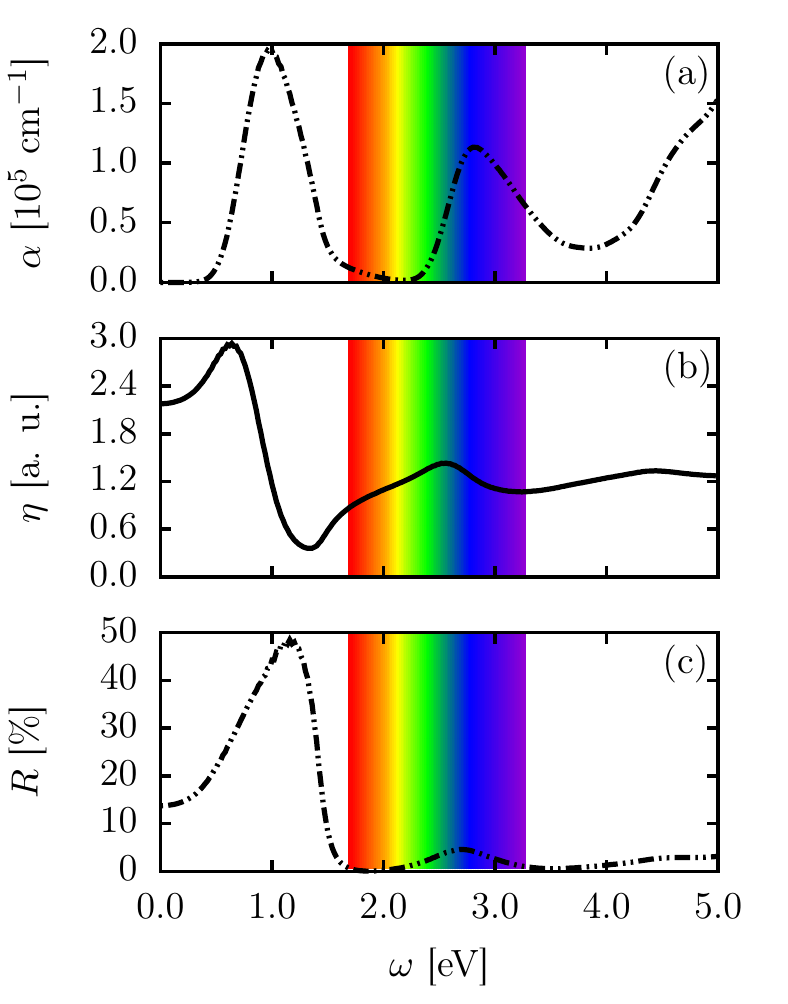}
	\caption{Irida-graphene absorption coefficient ($\alpha$), refractive ($\eta$), and reflectivity ($R$) indexes as a function of the photon energy.}
	\label{optical}
\end{figure}

Finally, we discuss the optical properties of IG. Figures \ref{optical}(a-c) show the absorption coefficient ($\alpha$), refractive ($\eta$), and reflectivity ($R$) indexes as a function of the photon energy value for the light propagation in the x-direction. Importantly, we did not notice a significant difference in the optical activity of the IG between the x- and y-directions. In this figure, one can note that the intense optical activity of IG is restricted to the infrared and violet regions, and a small activity in the ultra-violet region can be also noted.

The $\alpha$ profile presented in Figure \ref{optical}(a) revealed that the first optical transition peak occurs for nearly the same photon energy in all directions, about 1.0 eV (i.e., within the infrared spectrum). The second peak transition peak occurs within the violet region, about 3.0 eV. The maximum absorption intensities are approximately $2.0 \times 10^5$ cm$^{-1}$ and $1.2 \times 10^5$ cm$^{-1}$ within the infrared and violet regions, respectively. In the context of carbon-based materials, intense absorption within the infrared region denotes the presence of sp$^{2}$-like carbons \cite{rodil2005infrared}.
 
Figures \ref{optical}(b) and \ref{optical}(c) show the refractive and reflectivity indexes, respectively, as a function of the photon energy. The maximum peak for refractive index occurs at 0.7 eV, as depicted in Figure \ref{optical}(b). Light attenuation occurs for photon energies higher than 0.7 eV. $\eta$ values converge to 1.2, suggesting that the incident light tends to be refracted similarly in all directions. 

The reflective index has the maximum peak at 1.2 eV and drops to almost zero in the infrared region. A small reflection activity of IG, around 5\%, is note within the violet region. In this sense, all the incident violet and ultra-violet lights tends to be absorbed, as shown in Figure \ref{optical}(c). On the other hand, almost 50\% of the incident light is reflected in the infrared region. These results suggest that IG can act as violet collector for photon energies about 3.0 eV since it presents very low reflectivity and refractive index greater than one. 

\section{Conclusions}

In summary, we carried out DFT and reactive ReaxFF MD simulations to propose a new 2D all-sp$^2$ carbon allotrope named Irida-Graphene (IG) by employing a bottom-up approach. IG is composed of fused rings containing 3-6-8 carbon atoms. To examine the structural stability of IG, we calculated the phonon spectrum. AIMD simulations were also used to confirm its structural and dynamical stability.  

All calculated phonon frequencies are positive, thus confirming the stability of IG. Moreover, the highest phonon frequency is about 47 THz, similar to graphene and indicating the presence of only sp$^{2}$ carbon-carbon bonds. In the AIMD simulations, the total energy per atom for the IG lattice fluctuates around a steady level during the heating process. The lattice structure maintained its planarity with no bond reconstructions at the final stage of the simulation. These features suggest that IG also has good thermal stability.  

The estimated elastic moduli are $YM_{X}=80$ GPa and $YM_{Y}=113$ GPa. IG has clear in-plane anisotropic mechanical properties, as expected from its topology. The eight-membered rings of carbon atoms are responsible for the smaller mechanical resilience presented by IG compared with graphene (about 1 TPa).

In the calculated band structure, no band gap was found between the valence and conduction bands, indicating that IG is metallic. The DOS near the Fermi level is essentially formed by 2p$_z$ atomic orbitals, confirming its metallic signature. A Dirac cone is situated above the Fermi level and in the center of the band at the $Y$-point. 

The intense optical activity of IG is restricted to the infrared and violet regions. A small optical activity in the ultra-violet zone were also noted. In the context of carbon-based materials, intense infrared absorption denotes the presence of sp$^{2}$-like carbons. A small reflection activity of IG (around 5\%) is note in the violet zone. All the incident violet and ultra-violet lights tends to be absorbed. On the other hand, almost 50\% of the incident light is reflected in the infrared region. These results suggest that IG can act as violet collector for photon energies about 3.0 eV since it presents very low reflectivity and refractive index greater than one.

\section*{Acknowledgement}

The authors gratefully acknowledge the financial support from Brazilian research agencies CNPq and FAP-DF. L.A.R.J acknowledges the financial support from a Brazilian Research Council FAP-DF grants $00193-00000853/2021-28$ and CNPq grant $302236/2018-0$, respectively. L.A.R.J acknowledges CENAPAD-SP for providing the computational facilities. W.F.C and W.F.G acknowledge the financial support from FAPDF grants $00193-00000857/2021-14$ and  $00193-00000811/2021-97$, respectively. R.T.S.J. gratefully acknowledges CNPq (Grants 312180/2019-5 PQ-2 and 465741/2014-2 INCT on Cybersecurity). R.T.S.J. and L.A.R.J. gratefully acknowledge the support from ABIN grant 08/2019. M.L.P. J. and L.A.R.J. acknowledge N\'ucleo de Computaç\~ao de Alto Desempenho (NACAD) for providing the computational facilities through the Lobo Carneiro supercomputer. L.A.R.J. thanks Fundaç\~ao de Apoio \`a Pesquisa (FUNAPE), Edital 02/2022 - Formul\'ario de Inscriç\~ao N.4, for the financial support.

\bibliographystyle{plain}
\bibliography{bibliography.bib}
	
\end{document}